\documentclass[10pt,twocolumn,letterpaper]{article}

 \usepackage[pagenumbers]{cvpr} % To force page numbers, e.g. for an arXiv version

\usepackage{times}
\usepackage{multirow} 
\usepackage{epsfig}
\usepackage{graphicx}
\usepackage{amsmath}
\usepackage{amssymb}
\usepackage{subcaption}
\usepackage{overpic}
\usepackage{wrapfig}
\usepackage{booktabs}
\usepackage{verbatim}
\definecolor{cvprblue}{rgb}{0.21,0.49,0.74}
\usepackage[pagebackref,breaklinks,colorlinks,allcolors=cvprblue]{hyperref}

\def\paperID{8204} % *** Enter the Paper ID here
\def\confName{CVPR}
\def\confYear{2025}

\title{HUG: Hierarchical Urban Gaussian Splatting with Block-Based Reconstruction for Large-Scale Aerial Scenes}
\author{Mai Su\\
Peking University\\
Beijing, China\\
\and
Zhongtao Wang\\\
Peking University\\
Beijing, China\\
\and
Huishan Au\\
Peking University\\
Beijing, China\\
\and
Yilong Li\\
Peking University\\
Beijing, China\\
\and
Xizhe Cao\\
Peking University\\
Beijing, China\\
\and
Chengwei Pan\\
Beihang University\\
Beijing, China\\
\and
Yisong Chen\\
Peking University\\
Beijing, China\\
\and
Guoping Wang\\
Peking University\\
Beijing, China\\
}

\begin{document}
\twocolumn[{%
\renewcommand\twocolumn[1][]{#1}%
\maketitle
   %\fbox{\rule{0pt}{2in} \rule{\linewidth}{0pt}}
   \includegraphics[width=\linewidth]{fig/teaser.pdf}
% \vspace{-2em}
\captionof{figure}{Our method achieves the SOTA performance on the \textit{MatrixCity} dataset. In (a), we show the results of the best-performing baseline method, CityGS~\cite{liu2024citygaussian}. In (b), we present the results of our method, demonstrating clear improvements in visual quality. For reference, (c) displays the Ground Truth. The quantitative metrics for this view, shown in (a) and (b), underscore the superiority of our method in terms of reconstruction quality.
%textbf{(c)} shows the ground truth, providing a clear benchmark for evaluating the quality of the results. 
\vspace{1em}}
\label{fig:teaser}
}]

\begin{abstract}

3DGS is an emerging and increasingly popular technology in the field of novel view synthesis. Its highly realistic rendering quality and real-time rendering capabilities make it promising for various applications. However, when applied to large-scale aerial urban scenes, 3DGS methods suffer from issues such as excessive memory consumption, slow training times, prolonged partitioning processes, and significant degradation in rendering quality due to the increased data volume. To tackle these challenges, we introduce \textbf{HUG}, a novel approach that enhances data partitioning and reconstruction quality by leveraging a hierarchical neural Gaussian representation. We first propose a visibility-based data partitioning method that is simple yet highly efficient, significantly outperforming existing methods in speed. Then, we introduce a novel hierarchical weighted training approach, combined with other optimization strategies, to substantially improve reconstruction quality. Our method achieves state-of-the-art results on one synthetic dataset and four real-world datasets.

% not only the accuracy of the reconstruction needs to be considered, but also the capability of the system will be tested by the scale of the scene. 
\end{abstract}

\section{Introduction}
\label{sec:intro}
With the increasing complexity of large urban 3D scenes and the growing demand for high-quality rendering, efficient scene reconstruction and rendering techniques have become critical. In recent years, Radiance Field Rendering methods, particularly those based on Neural Radiance Fields~\cite{mildenhall2021nerf} , have gained significant attention and established themselves as the dominant paradigm for high-quality 3D scene reconstruction and rendering. More recently, 3D Gaussian Splatting~\cite{kerbl20233d} has emerged as a promising alternative, offering new possibilities for efficient scene representation. However, despite their effectiveness, these methods often face inefficiencies when dealing with large-scale scenes or intricate details.
To address these challenges, we propose a novel approach named HUG for large-scale urban scene reconstruction and rendering that optimizes both the data partitioning process and the reconstruction pipeline, while incorporating an efficient Level of Detail (LOD) representation. 

To reconstruct blocks of large urban scene, many existing methods train large redundant regions around the block boundaries to ensure boundary reconstruction accuracy, resulting in significant computational waste. In contrast, our approach begins with an enhanced block-based reconstruction pipeline that concentrates more on improving reconstruction quality within blocks. By reducing the need for redundant training regions and focusing more on the reconstruction quality within blocks, we release the waste of computing resources, particularly when processing complex or large-scale scenes, and enables improvements in reconstruction quality.

Building upon this foundation, we integrate a neural Gaussian representation with an efficient block-wise 
hierarchical LOD architecture. This combination ensures high-fidelity scene rendering with low computational cost, regardless of distance. By streamlining computational overhead and maintaining high rendering quality, we enable efficient representation of large-scale aerial scenes without compromising performance.

Finally, we demonstrate the effectiveness of our method by achieving State-of-the-Art (SOTA) results on five public benchmarks for large-scale aerial scenes.
%Our method not only excels in practical applications but also sets new standards on standardized testing platforms, showcasing its robustness and versatility in rendering complex environments.

Our contributions can be summarized as follows:
\begin{itemize}
\item We propose a simple and effective visibility-based partitioning strategy for large-scale aerial datasets, outperforms existing methods in speed, requiring about 1 minute to process each scene.
\item We introduce a hierarchical neural Gaussian representation tailored for block-based reconstruction of large-scale aerial urban scenes. Our work also presents a novel hierarchical weighted image supervision strategy along with additional optimizations to further enhance reconstruction quality.
\item Our method achieves state-of-the-art results on one synthetic and four real-world datasets, showcasing its advantage over existing approaches.
\end{itemize}

\section{Related Work}
\label{sec:related_work}

\subsection{Neural Rendering}

Neural rendering, particularly through Neural Radiance Fields (NeRFs)~\cite{mildenhall2021nerf, barron2022mipnerf360, barron2021mip, martin2021nerf, reiser2023merf, pumarola2021d, tancik2023nerfstudio, xu2022sinnerf}, has made significant strides in 3D scene reconstruction and novel view synthesis. NeRFs represent a scene as a continuous volumetric function, where a deep neural network is trained to map 3D coordinates and viewing directions to radiance and density values. This implicit representation, combined with volumetric rendering, yields highly realistic novel views. However, the method's dependence on dense sampling and prolonged network inference times often leads to high computational costs and limits its rendering quality. To address this, several approaches have sought to accelerate NeRF-based rendering. Techniques such as InstantNGP~\cite{muller2022instant} leverage multi-resolution hash grids and compact neural networks to achieve substantial speedups while maintaining high visual fidelity. Mip-NeRF 360~\cite{barron2022mipnerf360} extends NeRF's capabilities to unbounded scenes while maintaining excellent anti-aliasing and high-quality rendering by properly handling multi-scale observations. Similarly, Plenoxels~\cite{fridovich2022plenoxels} uses sparse voxel grids to efficiently represent the scene's continuous density field, delivering notable performance improvements.

\begin{figure*}[!ht]
\centering
 {\captionsetup{skip=-0.5pt}  % 仅对本图生效
   \includegraphics[width=1.0\linewidth]{fig/framework.pdf}
   %\fbox{\rule{0pt}{3in} \rule{.9\linewidth}{0pt}}

   \caption{An overview of our proposed method, HUG. \textbf{(a).} The sparse point cloud from COLMAP~\cite{schoenberger2016sfm} is used to uniformly partition the scene into blocks. \textbf{(b).} For each view, visibility masks for all blocks are generated by reprojecting 3D points within blocks into the image space and segmenting corresponding visible regions. \textbf{(c).} Views are assigned to training block $j$ if the visible sparse point cloud exceeds a threshold. \textbf{(d).} During training each block, anchors within the view frustum are selected based on their position and level. These selected anchors infer the neural Gaussians used for rendering and are optimized using our hierarchical weighted image supervision along with other constraints. \textbf{(e).} All trained blocks are seamlessly merged and rendered. }
   }
\label{fig:main}
\end{figure*}

Alongside NeRFs, point-based rendering has emerged as an alternative for faster scene rendering. 3D Gaussian Splatting (3DGS)~\cite{kerbl20233d} stands out by using 3D Gaussians as primitives, combining explicit geometry with rasterized rendering to enable real-time performance without sacrificing visual quality. Scaffold-GS~\cite{lu2023scaffold} introduces a hierarchical structure to further improve scene reconstruction. By organizing Gaussians into a scaffold-like framework, this approach enables more accurate and efficient scene representations. Several works have focused on optimizing and compressing Gaussian representations~\cite{fan2023lightgaussian, morgenstern2023compact}, offering inspiration for large-scale, high-quality scene reconstruction. Despite the promise of these works, challenges remain, particularly concerning storage cost when scaling to large scenes. 

To tackle the issue of resource saturation, we propose a divide-and-conquer data partitioning strategy that decomposes the environment into manageable regions, ensuring more efficient allocation of computational resources during training. Additionally, we implement a Level-of-Detail system that efficiently adapts the scene's complexity based on the viewer's perspective.

\subsection{Large Scale Scene Reconstruction}
Large-scale scene reconstruction is a challenging problem that involves generating high-fidelity 3D models from extensive data sources. Early approaches~\cite{fruh2004automated, li2008modeling, pollefeys2008detailed} restore camera poses using the Structure from Motion (SfM) method. Photo Tourism~\cite{snavely2006photo} and Building Rome in a Day~\cite{agarwal2011building} rely on large image datasets to generate precise 3D reconstructions. With the emergence of Neural Radiance Fields (NeRF)~\cite{mildenhall2021nerf}, the paradigm for large-scale scene reconstruction has shifted toward learning-based methods that synthesize photorealistic views of a scene from sparse or incomplete inputs. However, these techniques often face limitations in scaling to large, complex environments due to their high computational and storage demands. To address these issues, recent works such as BlockNeRF~\cite{tancik2022block} and MegaNeRF~\cite{turki2022mega} have employed divide-and-conquer strategies, partitioning scenes into smaller blocks and optimizing each independently, thereby reducing memory consumption. LesGo~\cite{cui2024letsgo} combined LiDAR point clouds with 3DGS, achieving impressive modeling and rendering quality for large-scale garages. Some recent Gaussian-based approaches~\cite{chen2024gigagsscalingplanarbased3d,lin2024vastgaussian,liu2024citygaussian} also deploy similar strategies to divide large scenes into small blocks and rebuild them respectively. However, these methods train redundant regions around the block boundaries to avoid boundary artifacts, leading to significant computational waste.
\par
Our method overcomes these inefficiencies by utilizing the visibility mask, allowing it to focus on the specific regions that require detailed reconstruction.
% correspondence between geometry and imagery, enabling a targeted reconstruction of the scene's critical regions. Rather than attempting to reconstruct large surrounding areas, our approach minimizes data redundancy by focusing on the specific regions that require detailed reconstruction. This strategy significantly reduces both computational overhead and memory consumption, offering a more efficient solution for large-scale scene reconstruction.

\subsection{Level of Detail Scene Representation}
The primary objective of Level-of-Detail (LOD) techniques is to balance computational efficiency with visual fidelity, reducing resource usage for distant or less critical objects while maintaining high-quality rendering. This is achieved by adjusting detail levels based on object proximity or relevance.
\par
In the context of learning scene representations, incorporating LOD has become a significant area of active research. Early efforts like Mip-NeRF~\cite{barron2021mip} and NGLOD~\cite{takikawa2021neural} have utilized hierarchical representations and multi-resolution voxel grids to represent scenes at varying levels of detail. 
Unlike NeRF-based methods, the LOD handling in 3DGS scene representation is more intuitive. Several Gaussian-based methods~\cite{lin2024vastgaussian,liu2024citygaussian} achieve the LOD structure by merging the properties of Gaussian volumes after training all the scenes.
\par
Inspired by Octree-GS~\cite{ren2024octreegsconsistentrealtimerendering}, we use a framework that combines a hierarchical octree structure with Scaffold-GS~\cite{lu2023scaffold}, our approach enables the simultaneous construction of Level-of-Detail (LOD) details and refinement of the scene representation during training, rather than deferring the generation of a compressed LOD model to later stages. 

\section{Method}

In this section, we introduce \textbf{HUG}, a \textbf{H}ierarchical \textbf{U}rban \textbf{G}aussian Splatting method designed for high-quality reconstruction and rendering of large-scale aerial urban scenes. An overview of the proposed method is illustrated in \Cref{fig:main}.

\subsection{Preliminary: 3D Gaussian Splatting}

3D Gaussian Splatting (3DGS)~\cite{kerbl20233d} represents a 3D scene using a set of anisotropic Gaussian primitives. Each Gaussian is defined by a mean position $\boldsymbol{\mu} \in \mathbb{R}^3$ and a covariance matrix $\boldsymbol{\Sigma} \in \mathbb{R}^{3 \times 3}$:
\begin{equation}
G(\boldsymbol{x}) = \exp\left( -\tfrac{1}{2} (\boldsymbol{x} - \boldsymbol{\mu})^\top \boldsymbol{\Sigma}^{-1} (\boldsymbol{x} - \boldsymbol{\mu}) \right),
\end{equation}
where $\boldsymbol{x} \in \mathbb{R}^3$ is a point in space. Each Gaussian also has an opacity $\alpha \in [0,1]$ and color features for view-dependent rendering. To synthesize images, Gaussians are projected into the image space using camera parameters, producing 2D Gaussian representations:
\begin{equation}
G'(\boldsymbol{x}') = \exp\left( -\tfrac{1}{2} (\boldsymbol{x}' - \boldsymbol{\mu}')^\top \boldsymbol{\Sigma}'^{-1} (\boldsymbol{x}' - \boldsymbol{\mu}') \right),
\end{equation}
where $\boldsymbol{x}' \in \mathbb{R}^2$ is a pixel position. The final image is rendered by compositing these projected Gaussians using alpha blending:
\begin{equation}
\boldsymbol{C}(\boldsymbol{x}') = \sum_{i} T_i\, \alpha_i\, G'_i(\boldsymbol{x}')\, \boldsymbol{c}_i, \quad T_i = \prod_{j=1}^{i-1} \left( 1 - \alpha_j\, G'_j(\boldsymbol{x}') \right),
\label{eq:gs_color}
\end{equation}
with $\boldsymbol{c}_i$ being the color of the $i$-th Gaussian. The rendering process is differentiable, allowing optimization of Gaussian parameters by minimizing the difference between rendered and ground truth images.

\subsection{Visibility-based Partitioning}
\label{sec:focused}
To reconstruct large-scale aerial urban scenes, the original 3DGS method must adopt a divide-and-conquer strategy due to memory limitations. We propose a simple yet efficient visibility-based partitioning strategy, which takes only 1 minute on the \textit{MatrixCity} dataset, whereas CityGS~\cite{liu2024citygaussian} requires approximately 2 hours. Additionally, we integrate an SVM mask derived from visibility, enabling our method to focus more effectively on in-block regions, further enhancing reconstruction quality.
% To facilitate efficient reconstruction of large-scale scenes while conserving computational resources, we propose a focused scene partition strategy. This approach partitions the scene into manageable blocks and selects relevant views for each block based on visibility. Our partition is simple yet effective, it only takes 1 minute for \textit{MatrixCity} scene while CityGS takes about two hours. By doing so, our method enhances reconstruction quality and minimizes computational redundancy.
\paragraph{Scene Partitioning.}
Our partitioning approach leverages the sparse point cloud $P$ generated by COLMAP~\cite{schoenberger2016sfm} as the foundation. The scene is uniformly divided into multiple blocks indexed by $j$, ensuring controlled block sizes to prevent memory overflow. Therefore, the initial sparse point cloud for block $j$ is $ P_j \subseteq P$.Each block will be reconstructed independently, and can be processed in parallel across multiple GPUs and machines, allowing for efficient use of computational resources.

\paragraph{Visibility-Based View Partitioning.}

To determine the appropriate training views for block $j$, we compute the total number of reprojected visible sparse points of block $j$ in image $I_i$. An image $I_i$ is included in the training set $\mathcal{T}_j$ for block $j$ if the number of visible points exceeds a threshold $\tau_{p}$:
\begin{equation}
I_i \in \mathcal{T}_j \quad \text{if} \quad \sum_{p \in P_j} \mathbb{I} \left( V(p, I_i) \right) > \tau_p
\end{equation}
where $\mathbb{I} \left( V(p, I_i) \right)$ is an indicator function that takes the value 1 if point $p \in P_j$ is visible in image $I_i$, and 0 otherwise. This selection criterion ensures that only images with sufficient visibility of block $j$ contribute to its reconstruction, reducing redundancy while preserving critical information.
% This ensures that only images that significantly cover block $j$ contribute to its reconstruction, thereby reducing computational waste and enhancing efficiency.
\paragraph{Visibility Mask Generation.}

For each image $I_i$ and block $j$, we generate a visibility mask $M_i^j(u,v)$ to indicate which regions of the image correspond to block $j$. The mask is computed by reprojecting the sparse 3D points of block $j$ into the image space, followed by non-linear classification to segment the visible region:
\begin{equation}
M_i^j(u,v) = 
\begin{cases}
1, & \text{if } (u,v) \text{ corresponds to \ block } j, \\
0, & \text{otherwise},
\end{cases}
\end{equation}
where $(u,v)$ are pixel coordinates in image $I_i$.

% \begin{figure}[t]
% \begin{center}
%    \includegraphics[width=1.0\linewidth]{fig/mask.pdf}
%    %\fbox{\rule{0pt}{3in} \rule{.9\linewidth}{0pt}}
% \end{center}
% \caption{Illustration of the importance of our visibility mask. \textbf{(a).} By applying the mask, we avoid reconstructing irrelevant regions outside the block, focusing computational resources on enhancing reconstruction quality within the block. \textbf{(b).} The upper figure shows the practices of most existing methods. When rebuilding a block, it is often accompanied by the reconstruction many outside areas, which will be cut during the fusion step. The lower figure shows our method, focusing more on the in-block area.} 
% \label{fig:mask}
% \end{figure}
The generated mask actively contributes to the training stage by concentrating on regions that provide key information for the target block, thereby reducing computational redundancy and enhancing reconstruction quality within the block.
% The mask obtained above is used not only in view selection but also in training stage, detailed in \Cref{sec_train}.  By concentrating on regions that provide crucial information about the target block. We mitigate the computational waste associated with processing Gaussians that may be discarded in later stages, as observed in other works such as CityGS~\cite{liu2024citygaussian} and VastGaussian~\cite{lin2024vastgaussian}.

\subsection{Hierarchical Neural Gaussian}
\label{sec_train}

We employ an Octree-based hierarchical reconstruction strategy, leveraging neural Gaussians derived from Anchors to optimize each block. Built upon Scaffold-GS~\cite{lu2023scaffold} and Octree-GS~\cite{ren2024octreegsconsistentrealtimerendering}, our system has a combination of LOD and Neural Gaussian capabilities. To the best of our knowledge, we are the first to apply a combination of block-trained neural Gaussians and LOD techniques throughout the entire pipeline to the reconstruction and rendering of aerial urban scenes. 
% Although GigaGS~\cite{chen2024gigagsscalingplanarbased3d} also use neural anchors during training, these anchors are discarded after training is completed.

\begin{figure}[t]
\begin{center}
   \includegraphics[width=1.0\linewidth]{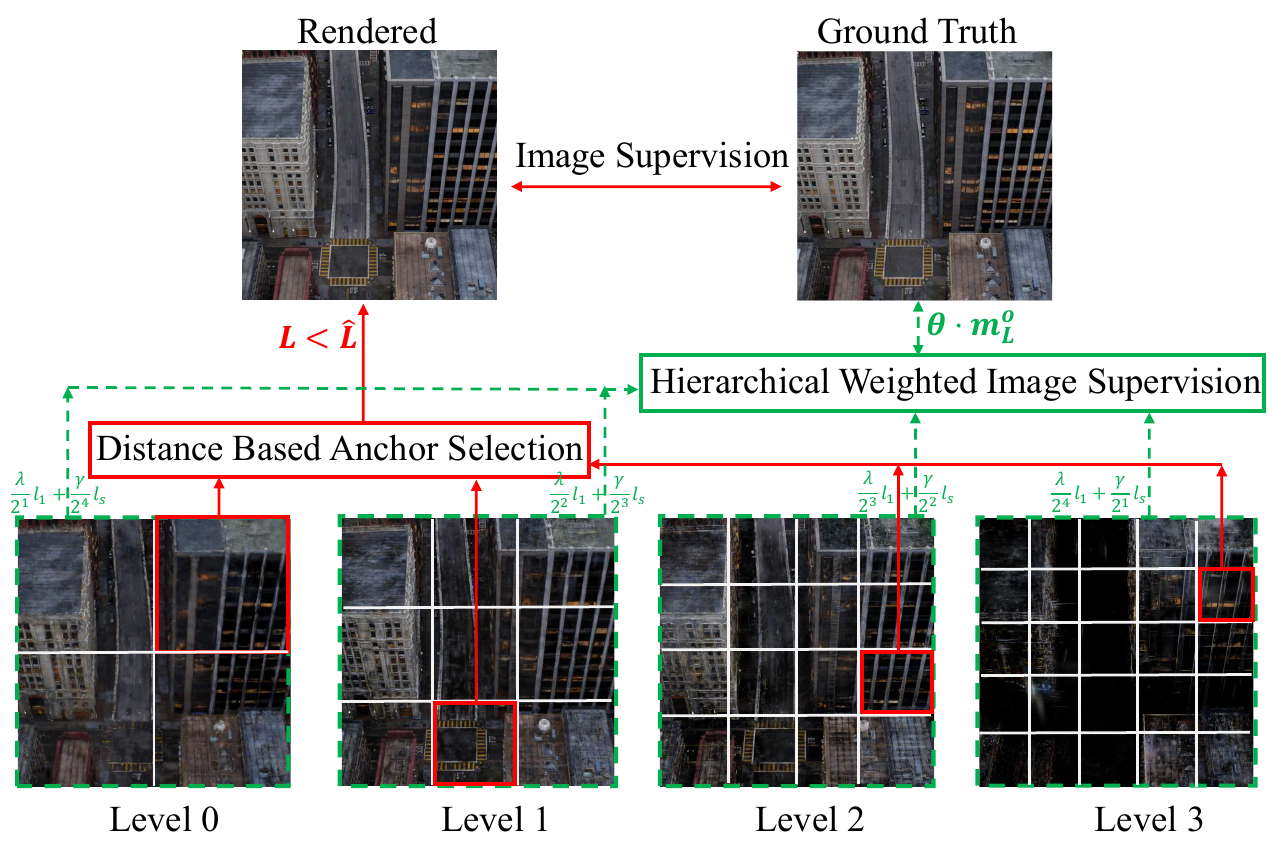}
   % \fbox{\rule{0pt}{2in} \rule{.9\linewidth}{0pt}}
\end{center}
   \caption{Hierarchical Octree Gaussian Training Strategy. We first select visible anchors based on the distance between the anchor and the camera center across all levels to render a complete scene image. Simultaneously, anchors from different levels are rendered separately, with distinct L1 and SSIM weights applied to evaluate the rendered image at each level.}
\label{fig:method}
\end{figure}

\paragraph{Hierarchical Anchor Initialization and Selection}
Our anchors are initialized at the centers of the nodes of an octree that covers the 3D space. The octree initialization for each block starts with a coarse 3D point cloud $P_j$ corresponding to block $j$, while the complete point cloud $P$ is typically obtained from COLMAP~\cite{schoenberger2016sfm} or PixSfM~\cite{lindenberger2021pixsfm}. Inspired by the Octree-GS~\cite{ren2024octreegsconsistentrealtimerendering}, the maximum level of the octree $K$ is determined by $d_{max}$ and $d_{min}$, the 0.95 and 0.05 quantiles of the distances between all points in $P_j$ and all the camera centers. It is calculated as follows:
\begin{equation}
\label{eq:lod_level}
    K= \lfloor log_{2}(d_{max}/d_{min}) \rceil + 1 
\end{equation}
where $(\lfloor \cdot \rceil)$ denotes the rounding operation. Anchors with various levels of detail (LOD) are generated in an octree manner, placed at the nodes that contain any point from $P_j$. For each training view, the predicted level for each visible anchor is calculated as follows:
% as described in Octree-GS~\cite{ren2024octreegsconsistentrealtimerendering}:
\begin{equation}
\label{eq:level_pred}
\hat{L} = \lfloor \min(\max(log_{2}(d_{max}/d), 0), K-1) \rfloor
\end{equation}
where $d$ is the distance between an anchor and the camera center. Only anchors with levels $L \leq  \hat{L}$ are selected for further optimization. As shown by the red-marked portion in \Cref{fig:method}. 
% During optimization, each selected anchor expands into 10 neural Gaussians, each with a explicit offset, neural scaling, rotation, color, and opacity. These neural Gaussians are rendered using differentiable rasterization as detailed in before works~\cite{kerbl20233d,ren2024octreegsconsistentrealtimerendering}.
% Before constructing the octree, we first divide the bounding space of $P$ into $2^{12}$ subspaces to better represent the scene and prevent the generation of large anchors. For each of the $2^{12}$ subspaces, starting from the coarsest level $0$ to the finest level $K-1$, it is subdivided into 8 equal parts at each level to construct the octree, retaining only nodes that contain points from $P$. The anchors are assigned to the centers of the nodes, with each anchor sharing the same level $L$ as its corresponding node.

\paragraph{Hierarchical Anchor Optimization} 

%\par 
%In Octree-GS~\cite{ren2024octreegsconsistentrealtimerendering}, a progressive training strategy is introduced, where the training iterations before a specified point are divided into several intervals. During the first interval, only anchors with levels ${L \leq  \lfloor K/2 \rfloor}$ can be selected by distance $d$. As the iterations progress, this constraint is gradually relaxed until anchors of all levels can be selected according to Equation \ref{eq:level_pred}. This progressive training and dynamic anchor selection strategy are adopted in our system. 
% \par
The selected anchors are first decomposed into multiple neural Gaussians, which are then rendered using alpha blending as defined in \Cref{eq:gs_color}. The resulting image is optimized against the training data using an L1 color loss $l_1$ and an SSIM structure loss $l_{s}$, following a similar approach to 3DGS~\cite{kerbl20233d} and Octree-GS~\cite{ren2024octreegsconsistentrealtimerendering}.
\par
% To enhance hierarchical capabilities and refine details at different levels, each training view will be rendered an additional \( K \) times with different LOD levels. 
In our work, we observe that optimizing hierarchical neural Gaussians using the aforementioned strategy leads to obvious floaters and suboptimal outcomes, as illustrated in \Cref{fig:anchor}. To resolve the problems above, a novel \textbf{hierarchical weighted image supervision} is introduced as shown by the green-marked portion of \Cref{fig:method}. 
\par
We begin by rendering anchors with levels $L \leq \hat{L}$ according to \Cref{eq:level_pred}, the rendered image $R_{full}$ is then compared with the ground truth image $gt$ using the L1 loss $l_1$ and SSIM loss $l_{s}$, weighted by ${\lambda}$ and ${\gamma=1-\lambda}$. Next, we render only the anchors with the finest level ${K-1}$, and this rendered image $R_{K-1}$ is evaluated with $l_1$ and $l_{s}$, but with weights ${\lambda/2^{K}}$ and ${\gamma/2^{1}}$, respectively. For anchors with levels equal to ${K-2}$, the weights are ${\lambda/2^{K-1}}$ and ${\gamma/2^{2}}$. This process continues until level 0.
\par
Our hierarchical weighted image supervision also incorporates an opacity mask $m^{o}_{L}$, which addresses the presence of extensive transparent regions that arise when rendering higher-level anchors, as shown by level 3 in \Cref{fig:method}. Additionally, a weighting factor $\theta$ is introduced to balance this component against the traditional training content. Accordingly, the hierarchical loss in our framework can be formulated as follows:
% \begin{equation}
% \label{eq:photo_loss}
% \begin{aligned}
% loss = \lambda \cdot l_1(gt, R_{full}) + \gamma\cdot l_{s}(gt, R_{full}) + \\
% \theta\cdot(\frac{\lambda}{2^{K}} \cdot l_1(gt, R_{K-1},m^{o}_{K-1})+ \frac{\gamma}{2^{1}} \cdot l_{s}(gt, R_{K-1},m^{o}_{K-1}) \\
% + \frac{\lambda}{2^{K-1}} \cdot l_1(gt, R_{K-2},m^{o}_{K-2}) + \gamma/2^{2} \cdot l_{s}(gt, R_{K-2},m^{o}_{K-2}) \\
% + ... +\lambda/2^{1} \cdot l_1(gt, R_{0}) + \gamma/2^{K-1} \cdot l_{s}(gt, R_{0}))
% \end{aligned}
% \end{equation}

% \begin{equation}
% \label{eq:photo_loss}
% \begin{aligned}
% \text{loss} = \lambda \cdot l_1(gt, R_{\text{full}}) + \gamma \cdot l_s(gt, R_{\text{full}}) \\
% + \theta \cdot \sum_{k=1}^{K} \left(  \frac{\lambda}{2^{K+1-k}} \cdot l_1(gt, R_{K-k}, m^o_{K-k}) \\
% + \frac{\gamma}{2^k} \cdot l_s(gt, R_{K-k}, m^o_{K-k}) \right)
% \end{aligned}
% \end{equation}

\begin{equation}
\label{eq:photo_loss}
\begin{aligned}
{l_{hier}} = & \; \lambda \cdot l_1(gt, R_{{full}})  + \gamma \cdot l_s(gt, R_{{full}}) \\
& + \theta \cdot \sum_{k=1}^{K} \left(  \frac{\lambda}{2^{K+1-k}} \cdot l_1(gt, R_{K-k}, m^o_{K-k}) \right. \\
&  \quad \left. + \frac{\gamma}{2^k} \cdot l_s(gt, R_{K-k}, m^o_{K-k}) \right)
\end{aligned}
\end{equation}

This approach is designed to render higher-level anchors with greater detail, while lower-level anchors focus on capturing approximate colors to serve as a foundation for the fully rendered images. This novel hierarchical weighted image supervision ensures that all anchors remain active during training, enhancing hierarchical optimization and improving overall image quality. Our hierarchical anchor supervision strategy effectively addresses the issue of anomalous rendering in OctreeGS~\cite{ren2024octreegsconsistentrealtimerendering}, as shown in \Cref{fig:anchor}.
% More details are shown in supplementary material.

\paragraph{Anchor Splitting}
\label{sec:split}
Splitting Gaussians or Anchors based on gradients, as proposed in 3DGS~\cite{kerbl20233d} and Scaffold-GS~\cite{lu2023scaffold}, involves dividing Gaussians with gradient values $\Delta g$ exceeding a threshold $\tau_g$ into two new Gaussians with similar properties.
\par
However, as training progresses, an increasing number of anchors tend to average the loss, causing the gradient threshold $\tau_g$ to become less effective. To address this, we propose an innovative \textbf{dynamic gradient threshold} strategy, where the gradient threshold for splitting is gradually decreased during training: $\tau_{g}(i)=\tau_{g}\cdot \eta^{\lfloor i/M \rfloor}$, where ${i}$ is the current iteration number, $M$ is the interval, and ${\eta}$ is the decrease rate. 
% To ensure adequate anchor splitting, we set a percentage threshold \( \tau_s \), and if the percentage of anchors meeting the criteria falls below this threshold, additional anchors with the highest $\Delta g$ values in the top $\tau_s$ are split.
% For anchors with particularly high gradients, i.e., ${\Delta g} > \beta \tau_{g}$ ($\beta > 1$), one generated anchor retains its original level $L$, while the other is assigned a higher level $L+1$, as described in Octree-GS~\cite{ren2024octreegsconsistentrealtimerendering}. 
% This operation can enhance the refinement of higher-level details. The splitting is discouraged if specific anchor points are located precisely at the target split node. However, our work found that this restriction degrades performance in detail-rich regions. To improve detail rendering in these areas, we removed this limitation. 
\par
In the training process, we observed that some anchors have gradients significantly greater than the gradient threshold $\tau_{g}(i)$, indicating that these anchors are unsuitable for remaining in the current LOD. To address this, we propose an innovative \textbf{dynamic anchor level transition} strategy that adaptively refines anchor levels as follow:
\begin{equation}
L = \min\left( K-1, L + \left\lfloor \sum(0.01\cdot\mathbb{I}(\Delta g > \beta \cdot \tau_g(i))) \right\rfloor \right)
\end{equation}
where $L$ is the initialized level of an anchor, $\beta$ is a hyperprameter larger than 1, and $\mathbb{I}$ is the indicator function that checks whether the anchor's gradient at a given training iteration $i$ surpasses $\beta$ times the dynamic gradient threshold. Each time the condition is met, the anchor's level $L$ is gradually increased by 0.01.
\par
This design allows anchors to dynamically adjust their levels based on local gradient magnitudes, ensuring higher-resolution details in regions where greater precision is needed. By progressively refining anchor levels, our approach enhances the adaptability of the hierarchical anchor representation, leading to improved reconstruction quality.

% ensuring higher-resolution details in regions where greater precision is needed.

% , where anchors with large gradients are moved to higher levels, allowing them to capture more detailed information. This adjustment ensures that anchors with higher levels are generated during training, enhancing the overall reconstruction quality.

% The Octree-GS~\cite{ren2024octreegsconsistentrealtimerendering} also gradually increase the level of each anchor with a gradient exceeding a threshold smaller than $\Delta_{g}$, by adding a property $L_{e}$. We adopt this strategy in our work as well. However, we found that during training, $L_{e}$ can exceed 1 or even reach values greater than 2, indicating that the total LOD level ${K}$, derived from \Cref{eq:lod_level}, may lack sufficient detail. Therefore, when the average value of the $L_{e}$ is larger than 1, we increase the LOD level of the scene by 1, and reset the $L_{e}$ to lower values.  This adjustment allows anchors with higher levels to be generated during the training process.
\paragraph{Anchor Pruning}
Pruning anchors or Gaussians based on opacity is a common technique. However, relying solely on opacity may lead to improper optimization of some anchors due to their LOD properties, as illustrated in \Cref{fig:anchor}. Previous works~\cite{ren2024octreegsconsistentrealtimerendering,lu2023scaffold,kerbl20233d} often overlooked Gaussians with low visibility counts during pruning. 
\par
In our work, we propose an innovative \textbf{visibility-based anchor pruning} strategy specially designed for hierarchical anchors. After every $M$ iterations of training with only distance-based anchor selection, anchors with a visibility count lower than ${\epsilon_{c}}$ will also be pruned.
% In addition to this universal pruning strategy, we introduce a visibility-based pruning approach. Since level-related anchor selection is applied in our work, new anchors occasionally appear at incorrect locations with incorrect levels. This results in anchors that retain only their initial properties and are rarely optimized in training views, yet they have to be rendered in test views if they are closer to the test view than to any training views, we show this in the supplementary material. To address this issue, we count the number of times an anchor is selected during training. If this count is below a threshold of ${\epsilon_{c}}$, the anchor is pruned. In addition, we track the accumulated gradient of each anchor and prune those with an accumulated gradient smaller than a threshold of ${\epsilon_{g}}$.

\subsection{Scene Fusion and Rendering}
\paragraph{Local Scene Refiltering}
After completing training in all blocks, previous works, such as CityGS~\cite{liu2024citygaussian}, typically refilter Gaussians whose center positions lie outside the block boundaries. However, we found that this approach can introduce artifacts near the boundaries. Specifically, even if an anchor's center is outside the block, the neural Gaussians it generates often have centers that fall within the block. 
\par
To address this, we propose an innovative \textbf{greedy vote-based anchor refiltering} strategy. For any given anchor, it is retained if its center lies within the block. If the anchor's center is outside the block but more than half of its generated neural Gaussians have centers within the block, the anchor is also preserved. Notably, our method follows a greedy approach, as it selectively retains anchors based on voting results but does not actively remove them.

% \begin{equation}
% \begin{cases}
% (C_a \notin B) \quad \text{and} \quad \frac{\left| \{ C_g \notin B : \, \forall C_g \in N_a \} \right|}{|N_a|} > 0.5,  \\
% (C_a \in B) \quad \text{and} \quad \frac{\left| \{ C_g \notin B : \, \forall C_g \in N_a \} \right|}{|N_a|} > 0.6, & 
% \end{cases}
% \end{equation}

% In this formulation, the anchor is refiltered out if either Condition 1 or Condition 2 is satisfied.
% We are the first to introduce neural octree Gaussians to urban scene reconstruction tasks. After training each block in parallel, we refilter the anchors within each block. We apply a a non-linear transformation similar to that used in CityGS~\cite{liu2024citygaussian} to map all anchor centers to the range [-2, 2]. The bounding box obtained from the block index is then used to refilter the anchors. Since each anchor has 10 neural Gaussians, and the offsets can sometimes be quite large, some anchors may be located inside the bounding box while their neural Gaussians are outside, or vice versa. Therefore, the positions of the neural Gaussians are also considered during the refiltering process. An anchor with more than 5 Gaussians inside the bounding box will be retained in the block, while anchors with fewer than 2 Gaussians inside the bounding box will be discarded, even if the anchor itself is within the block. Refiltered anchors of all blocks are then merged to form a global scene with an additional block index property. \par
\paragraph{Global Scene Rendering}
After refiltering, all blocks are stitched together. Methods like CityGS~\cite{liu2024citygaussian} can directly leverage the 3DGS~\cite{kerbl20233d} rasterization pipeline for global rendering. However, methods such as Octree-GS~\cite{ren2024octreegsconsistentrealtimerendering}, which incorporate neural Gaussians, rely on small MLPs. These MLPs are trained separately for different neural networks and thus have distinct weights, making Octree-GS unsuitable for direct application to multi-block scenes. To overcome this limitation, we reimplement the rasterization process to support the loading of multiple MLPs. During rendering, all visible anchors are first selected, and their corresponding neural Gaussians are then generated in parallel using the MLP trained on the respective block. This multi-MLP rendering strategy strikes a balance between leveraging the efficiency of MLPs and accommodating their limited learning capacity, enabling seamless rendering across blocks.
\begin{figure*}[!t] 
\centering
   \includegraphics[width=1.0\textwidth ]{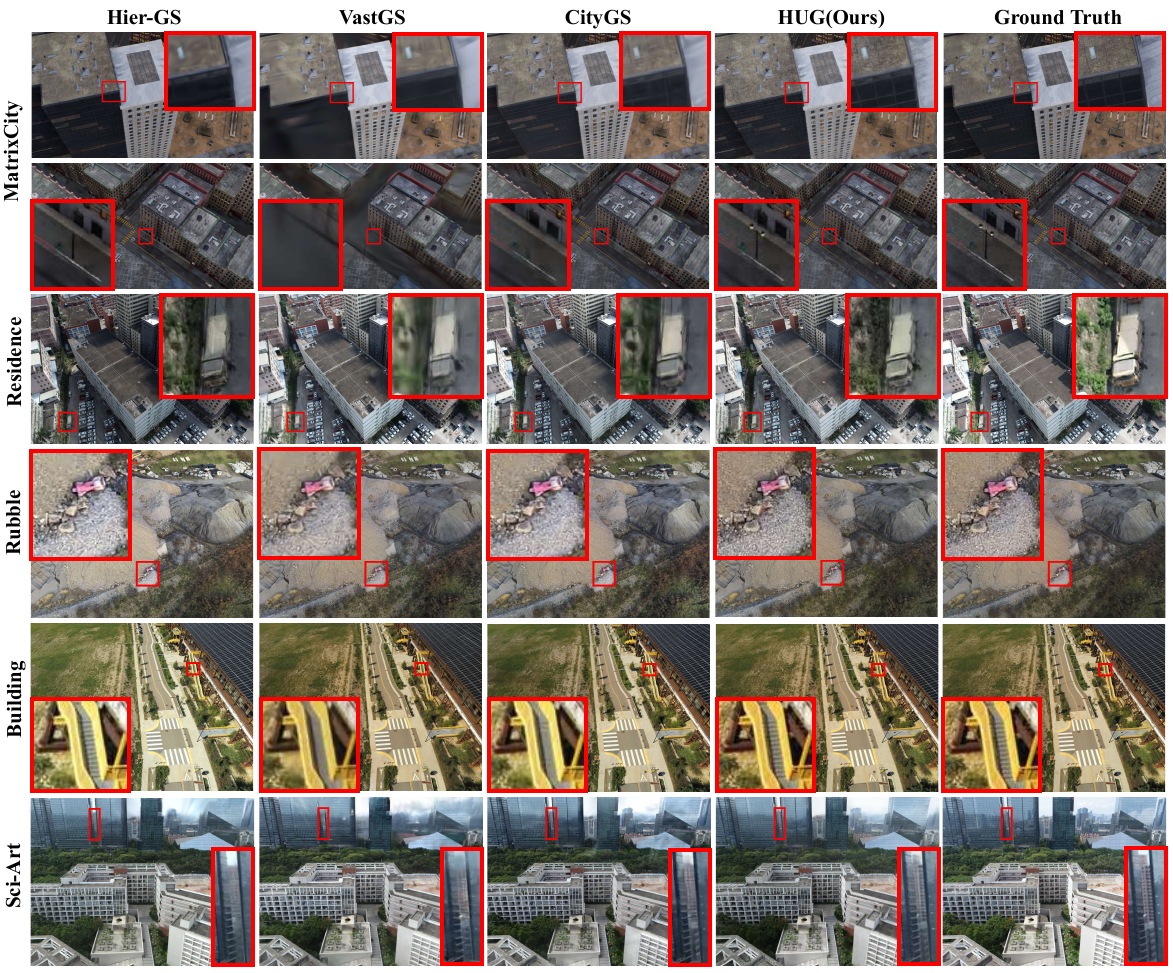}
   % \includegraphics[width=1.0\textwidth,trim={90 40 105 50},clip]{fig/fig_visual_compare.pdf}

   %\fbox{\rule{0pt}{3in} \rule{.9\linewidth}{0pt}}

%\fbox{\rule{0pt}{7in} \rule{.9\linewidth}{0pt}}
\caption{Qualitative comparison with state-of-the-art methods on one synthetic and four real-world datasets. \textcolor{red}{Red} insets highlight patches that reveal notable visual differences between these methods.} 
\label{result_fig} 
\vspace{-1em}
\end{figure*}

\section{Experiments}

\subsection{Implementation Details}
We built our method on the foundation of Octree-GS~\cite{ren2024octreegsconsistentrealtimerendering} for implementation. Our framework integrates innovative techniques, including visibility-based partitioning, hierarchical weighted image supervision, and a novel anchor splitting and pruning strategy. We modified the rendering code and the SIBR viewer~\cite{sibr2020} to support LOD rendering of neural Gaussians generated by multiple MLPs from different blocks, enabling real-time rendering of large-scale aerial urban scenes.
% All source code, datasets, and preprocessing scripts will be made publicly available upon publication.
\par
We first apply our proposed visibility-based partitioning method to divide the scene and views into multiple blocks. The point cloud threshold $\tau_p$ is setting as 800. We then perform block-wise scene optimization. The weights balancing components in the $l_{hier}$ are set as $\lambda=0.2$, $\gamma=0.8$, and $\theta=0.02$. For every $M=5000$ iterations training, anchors with visibility count lower than ${\epsilon_{c}}=5$ will be pruned, and gradient threshold $\epsilon_g=2e^{-6}$ will be decreased by $\eta=0.8$. The $\beta$ in dynamic anchor level transition is set as 4. And the opacity mask $m^o_{L}$ is obtained with pixels with opacity larger than 0.5. All experiments were conducted on a machine with 8 Nvidia A6000 48GB GPUs and 2 Intel Xeon 6330 CPUs.

\subsection{Evaluation}
% Unlike previous approaches focusing on partial areas~\cite{song2023city}, we reconstructed the entire city to assess our method's performance comprehensively. 
We evaluated our method on five diverse scenes spanning different scales and environments. Our experiments included a synthetic \textit{Small City} scene from the city-scale dataset \textit{MatrixCity}~\cite{li2023matrixcity}, covering an area of 2.7 $km^2$. Beyond synthetic data, we tested our approach on real-world datasets, including \textit{Residence}, \textit{Rubble}, \textit{Building}, and \textit{Sci-Art} from Mega-NeRF~\cite{turki2022mega}. Following prior works~\cite{turki2022mega, zhenxing2022switch, zhang2023efficient, liu2024citygaussian, chen2024gigagsscalingplanarbased3d, ren2024octreegsconsistentrealtimerendering, lu2023scaffold}, we downsampled the four real-world datasets by a factor of 4 and reduced the image resolution to 1.6k for \textit{MatrixCity}.
% We evaluated our method on 5 diverse scenes with varying scales and environments. We utilized a synthetic \textit{SMall City} scene from the city-scale dataset, \textit{MatrixCity}~\cite{li2023matrixcity}, which is 2.7 $km^2$ in size. In addition to synthetic data, we conducted experiments on real-world datasets, including \textit{Residence}, \textit{Rubble}, \textit{Building}, and \textit{Sci-Art} from Mega-NeRF~\cite{turki2022mega}. Consistent with prior work~\cite{turki2022mega, zhenxing2022switch, zhang2023efficient, liu2024citygaussian, chen2024gigagsscalingplanarbased3d, ren2024octreegsconsistentrealtimerendering, lu2023scaffold}, we reduced the image resolutions to 1.6k for \textit{MatrixCity} and down-sample the four real-world datesets by 4.
%To further demonstrate the generalization capabilities of our method, we also tested on the street-view scene \textit{Block_A} from \textit{MatrixCity}~\cite{li2023matrixcity}, which includes 4,076 training images and 495 test images.

For quantitative evaluation, we employed standard metrics: Peak Signal-to-Noise Ratio (\textbf{PSNR}), Structural Similarity Index Measure (\textbf{SSIM}), and Learned Perceptual Image Patch Similarity (\textbf{LPIPS})~\cite{zhang2018unreasonable}. 

%Rendering speed was assessed using frames per second (\textbf{FPS}) to gauge real-time performance. To ensure accurate timing measurements, all CUDA streams were synchronized before recording render times for each frame.

We compared our method against several state-of-the-art baselines, including Mega-NeRF~\cite{turki2022mega}, 3DGS~\cite{kerbl20233d}, CityGS~\cite{liu2024citygaussian}, OctreeGS~\cite{ren2024octreegsconsistentrealtimerendering}, VastGS~\cite{lin2024vastgaussian}, and Hier-GS~\cite{kerbl2024hierarchical}. More details on the settings for each method are provided in the supplement.

\subsection{Results Analysis}
% We present the quantitative and qualitative evaluations of our proposed method on several large-scale scene datasets. We conducted a fair comparison with several state-of-the-art methods. We also xxxx.
% \par

% \paragraph{Quantitative Comparisons} \Cref{tab:main_result} shows the mathematical results, except for the PSNR metric in the \textit{Sci-Art} scene, where our results are slightly lower than Mega-NeRF~\cite{turki2022mega}, our method outperforms all baseline methods across all metrics in every other scene. Our method outperform OctreeGS~\cite{ren2024octreegsconsistentrealtimerendering} over 1.5 PSNR, which shows the effectiveness of block-based reconstruction and other novelties in our method. We also overwhelm the CityGS~\cite{song2023city} and VastGS~\cite{lin2024vastgaussian}, this proves the effective of our combination of hierarchical representation and neural Gaussian. Our method achieves competitive results on 3 scenes from real world dataset. We also performs better than Hier-GS~\cite{kerbl2024hierarchical}, which we believe is because this work is specially designed for large-scale street view scenes. All these results also prove the effectiveness of our visibility-based partitioning.

\paragraph{Quantitative Comparisons} \Cref{tab:main_result} presents the quantitative results. Except for the PSNR metric in the \textit{Sci-Art} scene, possibly due to the large number of distant views, where our results are lower than Mega-NeRF~\cite{turki2022mega} but still outperform all GS-based methods. Across all other metrics, it surpasses all baseline methods. Notably, our method exceeds Octree-GS~\cite{ren2024octreegsconsistentrealtimerendering} by over 1.5 PSNR, demonstrating the effectiveness of our block-based reconstruction and other proposed innovations. Additionally, we achieve superior performance compared to CityGS~\cite{song2023city} and VastGS~\cite{lin2024vastgaussian}, further validating the effectiveness of our hierarchical representation combined with neural Gaussians. Our method also outperforms Hier-GS~\cite{kerbl2024hierarchical}, which we attribute to our specific design tailored for large-scale aerial scenarios. These results further underscore the efficacy of our visibility-based partitioning strategy.
% our method achieves SOTA performance compared with novel NeRF-based and Gaussian-based methods on the MatrixCity~\cite{li2023matrixcity} dataset.
% Our method is not as good as VastGS, which we think is mainly because they adopt a decoupled appearance modeling, this is important for real world datasets like Mill-19, because the brightness and exposure tend to be are usually variable in these datasets. 

% consistently outperforms the baselines across all key metrics. These results demonstrate the effectiveness of our approach in achieving higher reconstruction quality, accurately capturing fine details and textures in complex scenes. 
%Additionally, our method maintains efficient rendering performance, highlighting the advantages of integrating an octree-based Level-of-Detail framework with 3D Gaussian Splatting for large-scale scene reconstruction and real-time rendering.

\begin{table*}[!ht]
  \centering
  \tabcolsep=0.1cm
  \resizebox{0.95\textwidth}{!}{%
 \begin{tabular}{@{}l|lll|lll|lll|lll|lll@{}}
  
    \toprule
      & \multicolumn{3}{c|}{MatrixCity} & \multicolumn{3}{c|}{Residence} & \multicolumn{3}{c|}{Rubble} & \multicolumn{3}{c}{Building} & \multicolumn{3}{c}{Sci-Art} \\
    \midrule
    Metrics  & SSIM$\uparrow$ & PSNR$\uparrow$ & LPIPS$\downarrow$  & SSIM$\uparrow$ & PSNR$\uparrow$ & LPIPS$\downarrow$ & SSIM$\uparrow$ & PSNR$\uparrow$ & LPIPS$\downarrow$ & SSIM$\uparrow$ & PSNR$\uparrow$ & LPIPS$\downarrow$  & SSIM$\uparrow$ & PSNR$\uparrow$ & LPIPS$\downarrow$ \\
    \midrule
    MegaNeRF~\cite{turki2022mega} & - & - & - & 0.628 & \cellcolor{orange!25}22.08 & 0.489 & 0.553 & 24.06 & 0.516 & 0.547 & 20.93 & 0.504 & 0.770  &\cellcolor{red!25}25.60 &0.390 \\

    3DGS~\cite{kerbl20233d} & 0.735 & 23.67 & 0.384 & \cellcolor{orange!25}0.791 & 21.44 & 0.236 & \cellcolor{yellow!25}0.777 & \cellcolor{yellow!25}25.47 & 0.277 & 0.720 & 20.46 & 0.305 & \cellcolor{yellow!25}0.830 &21.05 &0.242\\

    Octree-GS~\cite{ren2024octreegsconsistentrealtimerendering} & 0.814 & 26.41 & 0.282  & - & - & - & - & - & - & - & - & - & - & - & - \\

    Hier-GS~\cite{kerbl2024hierarchical} &\cellcolor{yellow!25}0.842 &\cellcolor{yellow!25}26.67 &\cellcolor{yellow!25}0.251 &0.776 &20.24 &\cellcolor{yellow!25}0.221 &0.765 &22.28 &\cellcolor{yellow!25}0.257 &0.733 &20.04 &\cellcolor{yellow!25}0.262 &0.828 &19.74 &\cellcolor{orange!25}0.207\\

    VastGS~\cite{lin2024vastgaussian} &- &- &- &\cellcolor{yellow!25}0.777 &20.38
  & 0.247  &0.766 &25.10 &0.294 &\cellcolor{yellow!25}0.740 &\cellcolor{orange!25}21.69 &0.293 &0.809 &\cellcolor{yellow!25}21.66 &0.269 \\

    CityGS~\cite{liu2024citygaussian} & \cellcolor{orange!25}0.865 & \cellcolor{orange!25}27.46 & \cellcolor{orange!25}0.204  & \cellcolor{red!25}0.813 & \cellcolor{yellow!25}22.00 & \cellcolor{orange!25}0.211 & \cellcolor{orange!25}0.813 & \cellcolor{orange!25}25.77 & \cellcolor{orange!25}0.228 & \cellcolor{orange!25}0.778 & \cellcolor{yellow!25}21.55 & \cellcolor{orange!25}0.246 &\cellcolor{orange!25}0.837 &21.39 &\cellcolor{yellow!25}0.230\\

    Ours & \cellcolor{red!25}0.883 & \cellcolor{red!25}28.02 & \cellcolor{red!25}0.142 & \cellcolor{red!25}0.813 & \cellcolor{red!25}22.33 & \cellcolor{red!25}0.207 & \cellcolor{red!25}0.839 & \cellcolor{red!25}26.42 & \cellcolor{red!25}0.197 & \cellcolor{red!25}0.792 & \cellcolor{red!25}22.35 & \cellcolor{red!25}0.228 &\cellcolor{red!25}0.846 &\cellcolor{orange!25}21.83 &\cellcolor{red!25}0.204\\
    \bottomrule
  \end{tabular}

  }
    \caption{Quantitative comparison of our method with state-of-the-art methods. The best scores in each column are highlighted with \textcolor{red}{red} background, second best with \textcolor{orange}{orange}, and third best with \textcolor{yellow}{yellow}. The symbol `-' denotes experiments that are difficult to reproduce.
  }
  \label{tab:main_result}
\end{table*}

\paragraph{Qualitative Comparisons}
\Cref{result_fig} provides qualitative comparisons between our method and others across five diverse scenes. In the \textit{MatrixCity} scene, our method excels at reconstructing intricate details, such as rooftop textures, facade steel structures, and street lamps, while preserving the fine structures with high fidelity. In the \textit{Residence} scene, our approach captures the truck's details more thoroughly than the baseline methods. In the \textit{Rubble} scene, while all methods successfully reconstruct the red object, our method stands out by exhibiting fewer artifacts in the surrounding rubble piles. In the \textit{Building} scene, both our method and Hier-GS~\cite{kerbl2024hierarchical} reconstruct the staircase steps accurately, but Hier-GS~\cite{kerbl2024hierarchical} suffers from noticeable color deviations. Finally, in the \textit{Sci-Art} scene, our method successfully recovers a building nestled between two background structures, which other methods fail to reconstruct entirely.
\par
\begin{figure}[!h]
% \captionsetup{skip=pt} % 让标题和图表更紧凑
% \vspace{-4pt}  % 减少图像与下方文字的间距
    \centering
    \includegraphics[width=0.99\linewidth]{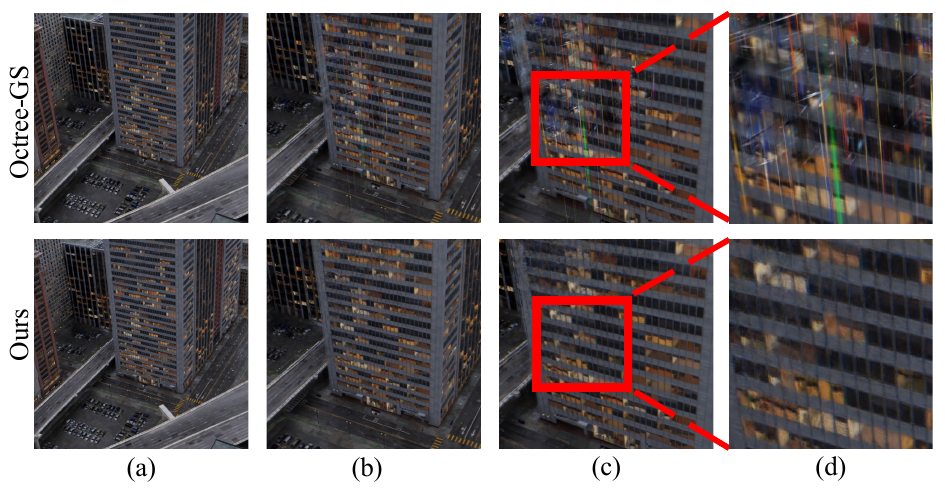}
    \caption{The unoptimized anchor issue of Octree-GS~\cite{ren2024octreegsconsistentrealtimerendering}. \textbf{(a).} A training view. \textbf{(b).} As the perspective zooms in, multicolored artifacts begin to appear in Octree-GS~\cite{ren2024octreegsconsistentrealtimerendering}. \textbf{(c).} Further zooming amplifies these artifacts, whereas our method effectively eliminates them. \textbf{(d)}. Close-up view of the issue.}
    \label{fig:anchor}
    % \vspace{-16pt}  % 减少图像与下方文字的间距
\end{figure}

\Cref{fig:anchor} also illustrates how our method effectively eliminates the unoptimized anchor issue in Octree-GS~\cite{ren2024octreegsconsistentrealtimerendering}, which often becomes apparent when zooming in on the scene.
% Our method produces sharper and more detailed reconstructions, accurately capturing fine textures and structural details with fewer artifacts.

% In particular, our reconstructions exhibit better preservation of intricate features such as building facades, textures, and edges. The car we rendered are even clearer than the ground truth. We better restored the lighting intensity of the lawn, showing the effectiveness of our weighted hierarchical supervision. The visual fidelity is consistent across different scenes, highlighting the robustness of our approach.

\paragraph{Partitioning and Optimization Efficiency}  
Partitioning large-scale scenes into smaller blocks followed by optimization has become a standard practice. Since these two stages are executed sequentially, the time consumption of each directly impacts the overall performance. \Cref{tab:performance_comparison} presents the time consumption of our method compared to related approaches. Our visibility-based partitioning is both simple and efficient, consistently completing in about one minute, whereas CityGS~\cite{liu2024citygaussian} requires over two hours for the \textit{MatrixCity} dataset. Efficient optimization within each block is equally crucial. While most methods, including ours, maintain an optimization time of approximately one hour per block, Hier-GS~\cite{kerbl2024hierarchical} takes nearly two hours due to additional post-processing refinements applied to each block.

\begin{table}[!t]
    \centering
     \resizebox{0.45\textwidth}{!}{
    \begin{tabular}{c cc cc cc cc cc}
        \toprule
        Scene & \multicolumn{2}{c}{MatrixCity} & \multicolumn{2}{c}{Rubble} & \multicolumn{2}{c}{Residence} & \multicolumn{2}{c}{Building} & \multicolumn{2}{c}{Sci-Art} \\
        \midrule
        Task & Part. & Opt. &  Part. & Opt. &  Part. & Opt. &  Part. & Opt. &  Part. & Opt. \\
        \midrule
        CityGS~\cite{liu2024citygaussian} & 188 & 61 & 77 & 51 & 118 & 79 & 97 & 78 & 71 & 55 \\
        VastGS~\cite{lin2024vastgaussian} & 51 & 89 & 4 & 44 & 10 & 59 & 8 & 68 & 20 & 52 \\
        Hier-GS~\cite{kerbl2024hierarchical} & 9.8 & 119 & 12.7 & 101 & 8.0 & 137 & 4.4 & 130 & 7.6 & 120 \\
        Ours & 1 & 67 & \textless1 & 52 & \textless1 & 78 & \textless1 & 63 & \textless1 & 59 \\
        \bottomrule
    \end{tabular}}
    \caption{Partitioning and optimization efficiency of different methods across various scenes, with time measured in minutes.}
    \label{tab:performance_comparison}
\end{table}

% We are the first to introduce a hierarchical neural Gaussian pipeline to urban scene reconstruction task, our main purpose is to allow partitioned parallel training and render the whole urban scene with low memory consumption, we compare with CityGS~\cite{liu2024citygaussian} both in LOD mode and non-LOD mode. The CityGS-LOD mode uses Gaussian compression method LightGaussian~\cite{fan2023lightgaussian} to compress Gaussians and reduce computational cost. We render \textit{Rubble} scene with similar visual quality while gradually raising the view center. The results are shown in \Cref{fig:efficiency}. As the view height increases, both CityGS and CityGS-LOD have a increase in number of Guassians. But, with our hierarchical representation, the number of Gaussians initially increases and then decreases. Our rendering speed outperforms CityGS, though it is slightly slower than CityGS-LOD. We believe this discrepancy is due to differences in compression rates, and our method involves selecting anchors and predicting neural properties. 
 
% \begin{figure}[!h]
% \begin{center}
%    \includegraphics[width=1.0\linewidth]{fig/efficiency compare.pdf}
%    %\fbox{\rule{0pt}{3in} \rule{.9\linewidth}{0pt}}
% \end{center}
% \caption{Rendered Gaussian count and FPS performance tested by increasing the camera height in the $Rubble$ scene.} 
% \label{fig:efficiency}
% \end{figure}

\begin{table}[!h] 
\centering 
\resizebox{0.75\linewidth}{!}{
\begin{tabular}{lccc} 
\toprule 
Model Setting & SSIM $\uparrow$ & PSNR $\uparrow$ & LPIPS $\downarrow$ \\
\midrule Baseline & 0.747 & 25.06 & 0.281 \\
add C1 & 0.779 & 25.40 & 0.244 \\
add C2 & 0.807 & 25.74 & 0.229 \\
add C3 & 0.834 & 26.33 & 0.201 \\
Full model & 0.839 &26.42 &0.197 \\
\bottomrule
\end{tabular}} 
\caption{Ablation results on \textit{Rubble} scene.} 
\label{tab_mask} 
\vspace{-1em}
\end{table}

\paragraph{Ablation}
To comprehensively assess the impact of each component in our proposed method, we conducted an ablation study on the \textit{Rubble} scene. Specifically, we evaluate the effectiveness of our approach by progressively incorporating key components: visibility-based masking (C1), hierarchical weighted image supervision (C2), dynamic gradient thresholding (C3), and finally, our full model. \Cref{tab_mask} presents the corresponding results, demonstrating the effectiveness of our proposed strategies.

\section{Conclusion}

In this paper, we present \textbf{HUG}, a novel Hierarchical Urban Gaussian Splatting method designed to reconstruct and render large-scale aerial urban scenes with high quality. Comprehensive experimental results demonstrating State-of-the-Art performance of our model, proves the effectiveness of our proposed visibility-based partitioning, hierarchical weighted image supervision, dynamic gradient threshold, and other proposals.

% , validate the effectiveness and robustness of our approach. Our method not only improves 

Despite our approach's advantages, a few limitations should be addressed in future work. First, our method does not account for dynamic objects. This can be improved by incorporating lighting decoupling and appearance encoding. Second, the data partitioning process relies on the sparse point cloud generated by COLMAP~\cite{schoenberger2016sfm}, which may lead to issues in low-texture areas. Combining pose-free Gaussian methods~\cite{hong2024pf3plat} may help it. Finally, our method does not support integrated aerial-ground training and rendering. This functionality could be incorporated in future research to enable more comprehensive scene reconstruction.

{
    \small
    \bibliographystyle{ieeenat_fullname}
    \bibliography{main}
}
\clearpage
\setcounter{page}{1}

%% TITLE
\maketitlesupplementary
We provide a more detailed discussion on the settings of the compared methods, view partition results, and additional qualitative comparisons. A rendered video is also included for a more comprehensive comparison.
\section{Settings of Compared Methods}
We compared our method against several state-of-the-art baselines, including Mega-NeRF~\cite{turki2022mega}, 3DGS~\cite{kerbl20233d}, CityGS~\cite{liu2024citygaussian}, OctreeGS~\cite{ren2024octreegsconsistentrealtimerendering}, VastGS~\cite{lin2024vastgaussian}, and Hier-GS~\cite{kerbl2024hierarchical}. The quantitative results for Mega-NeRF~\cite{turki2022mega}, 3DGS~\cite{kerbl20233d}, and CityGS~\cite{liu2024citygaussian} are taken from the CityGS~\cite{liu2024citygaussian} paper. The quantitative results for OctreeGS~\cite{ren2024octreegsconsistentrealtimerendering} are sourced from the OctreeGS~\cite{ren2024octreegsconsistentrealtimerendering} paper. The results for Hier-GS~\cite{kerbl2024hierarchical} were obtained using the released source code with all default settings. The results for VastGS~\cite{lin2024vastgaussian} are derived from the unofficial code available at https://github.com/kangpeilun/VastGaussian, with the same block number as CityGS~\cite{liu2024citygaussian} and ours. The qualitative results for CityGS~\cite{liu2024citygaussian} were obtained from the checkpoint provided by the authors.

\section{Data Partitioning Results}

\begin{figure}[!h]
\begin{center}
   \includegraphics[width=1.0\linewidth]{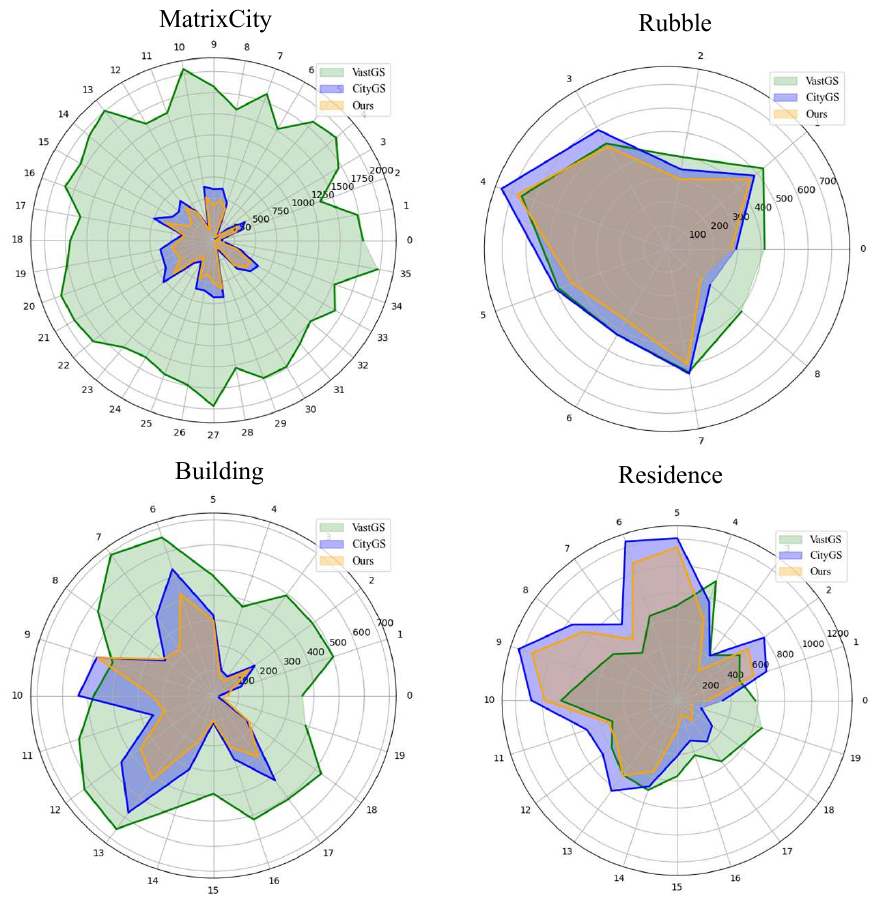}
   %\fbox{\rule{0pt}{3in} \rule{.9\linewidth}{0pt}}
\end{center}
\caption{Comparison of view partition results among our method, VastGS~\cite{lin2024vastgaussian}, and CityGS~\cite{liu2024citygaussian} across four datasets. The radial coordinate indicates the number of training images assigned to each block, with each vertex representing a specific block. As shown, our method consistently assigns fewer images on average, demonstrating a more efficient partition strategy.} 
\label{fig:partition_compare}
\end{figure}

To associate all training images with their respective blocks, we compute the total number of visible sparse points in each view. A threshold of \(\tau_p = 800\) is then applied to determine the assignment of each training image to specific blocks. In CityGS~\cite{liu2024citygaussian}, a coarse global 3DGS~\cite{kerbl20233d} model must first be constructed, followed by multiple renderings of all images corresponding to the number of blocks. VastGS~\cite{lin2024vastgaussian} introduces a more complex camera selection strategy, incorporating position-based data selection, visibility-based camera selection, and coverage-based point selection, resulting in a significantly higher number of assigned cameras per block. VastGS\cite{lin2024vastgaussian} has an average of over 1,000 views per block on the MatrixCity dataset, which exceeds GPU memory limits for a single block. This is why we do not report its numerical results on this scene. For visual comparisons, we avoid that specific block whenever possible. In contrast, our data partitioning strategy assigns fewer images to each block, reducing memory consumption during training and allowing more computational resources to be dedicated to refining the in-block reconstruction. Figure \ref{fig:partition_compare} visualizes the number of training images per block, illustrating that our method allocates fewer images on average while still achieving superior rendering performance, as demonstrated in Table \ref{tab:main_result}.
% To associate all training images with respective blocks, we calculate the total visible sparse point cloud in each view space. A threshold of $\tau_p = 800$ is then applied to assign each training image to specific blocks. In CityGS~\cite{liu2024citygaussian}, a coarse global 3DGS~\cite{kerbl20233d} model must first be constructed, after which all images are rendered multiple times corresponding to the number of blocks. In VastGS, a more complex camera selection is proposed consisting of position-based data selection, visibility-based camera selection, and coverage-based point selection, which leads to lots of cameras assigned to each block. Our data partitioning strategy assigns fewer images to each block, reducing memory requirements during training and allowing more resources to be dedicated to reconstructing the in-block area. The number of training images per block is visualized in Figure \ref{fig:partition_compare}. We can see that we allocate fewer images to each block on average, while still achieving better rendering performance as demonstrated in Table \ref{tab:main_result}.

\begin{figure*}[!h]
        \centering
    \includegraphics[width=0.95\linewidth]{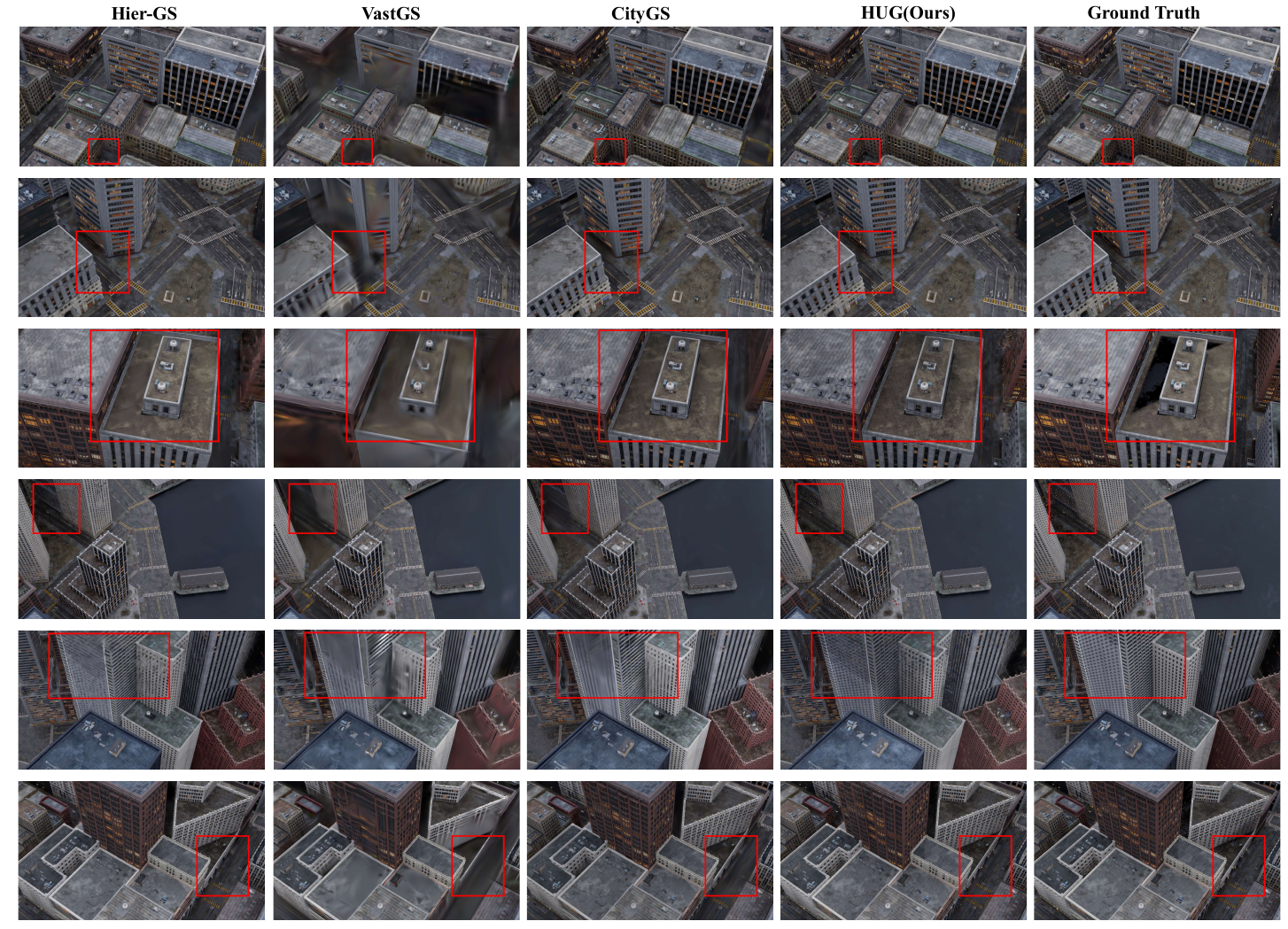}
    \caption{More qualitative comparisons on \textit{MatrixCity} dataset.}
    \label{fig:more_mccity}
\end{figure*}
\begin{figure*}[!ht]
    \centering
    \includegraphics[width=0.95\linewidth]{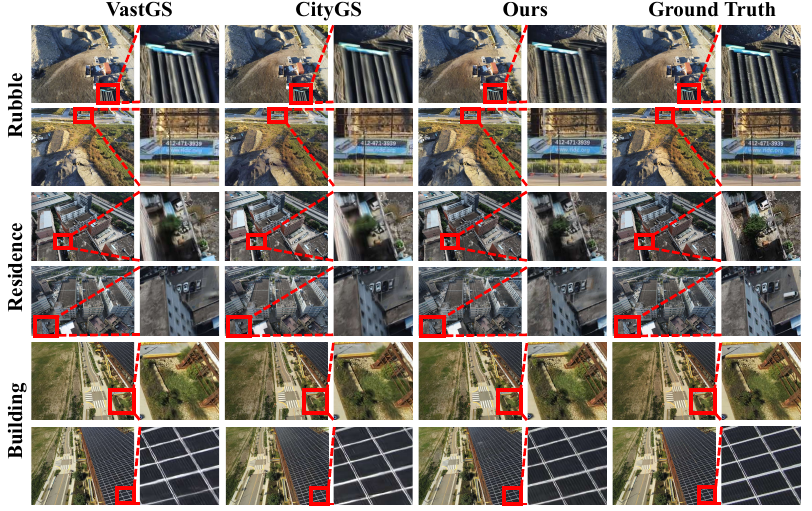}
    \caption{More qualitative comparisons on \textit{Rubble}, \textit{Residence} and \textit{Building} datasets.}
    \label{fig:more_compare}
\end{figure*}

\section{More Qualitative Comparisons}

We provide additional qualitative comparisons in Figure \ref{fig:more_mccity} and Figure \ref{fig:more_compare}.

\end{document}